\documentclass[aps,reprint,superscriptaddress,nofootinbib,amsmath,amssymb,physrev,nolongbibliography]{revtex4-2}

\newcommand{\ZIB}{Zuse Institute Berlin, Takustraße 7, 14195 Berlin, Germany}
\newcommand{\JCM}{JCMwave GmbH, Bolivarallee 22, 14050 Berlin, Germany}
\newcommand{\KIT}{Institute of Nanotechnology, Karlsruhe Institute of Technology, Kaiserstrasse 12, 76131 Karlsruhe, Germany}
\newcommand{\KITB}{Institute of Theoretical Solid State Physics, Karlsruhe Institute of Technology, Kaiserstrasse 12, 76131 Karlsruhe, Germany}
\newcommand{\AMOLF}{Department of Information in Matter and Center for Nanophotonics, AMOLF, Science Park 104, NL1098XG Amsterdam, The Netherlands}
\newcommand{\ARCNL}{Advanced Research Center for Nanolithography, Science Park 106, 1098 XG Amsterdam, The Netherlands}

\usepackage{siunitx}
\usepackage{graphicx}
\usepackage{xcolor,colortbl}
\usepackage[T1]{fontenc}

\usepackage[colorlinks=true, linkcolor=blue, citecolor=blue, urlcolor=blue]{hyperref} 
\begin{document}
\title{Uncovering Hidden Resonances in Non-Hermitian Systems with Scattering Thresholds}

\author{Fridtjof Betz}
\affiliation{\ZIB}
\author{Felix Binkowski}
\affiliation{\ZIB}
\affiliation{\JCM}
\author{Jan David Fischbach}
\affiliation{\KIT}
\author{Nick Feldman}
\affiliation{\ARCNL}
\affiliation{\AMOLF}
\author{Lin~Zschiedrich}
\affiliation{\JCM}
\author{Carsten Rockstuhl}
\affiliation{\KIT}
\affiliation{\KITB}
\author{A. Femius Koenderink}
\affiliation{\AMOLF}
\author{Sven Burger}
\affiliation{\ZIB}
\affiliation{\JCM}

\begin{abstract}
The points where diffraction orders emerge or vanish in the propagating spectrum of periodic non-Hermitian systems are referred to as scattering thresholds. Close to these branch points, resonances from different Riemann sheets can tremendously impact the optical response. However, these resonances are so far elusive for two reasons. First, their contribution to the signal is partially obscured, and second, they are inaccessible for standard computational methods. Here, the interplay of scattering thresholds with resonances is explored and a multi-valued rational approximation is introduced to access the hidden resonances. The theoretical and numerical approach is used to analyze the resonances of a plasmonic line grating. This work elegantly explains the occurrence of pronounced spectral features at scattering thresholds applicable to many nanophotonic systems of contemporary and future interest.
\end{abstract}

\maketitle

\section{Introduction}

In the early years of the twentieth century, Wood was intrigued by sudden changes in the intensity of light diffracted from a grating. He noticed that, under certain conditions, a small change in the wavelength could drop the intensity from maximum to minimum~\cite{Wood1902}. Wood described these features of the spectrum as the most interesting problem he had encountered so far and termed them anomalies, since they could not be explained by ordinary grating theory. A few years later, in a note on this phenomenon, Lord Rayleigh came to the conclusion that anomalies were to be expected at the scattering thresholds, that is, whenever a higher diffraction order appears in the propagating spectrum~\cite{rayleigh1907}. In 1941, Fano attributed another type of Wood's anomalies to surface plasmons, and thus established a connection to resonances~\cite{Fano1941}. Eventually, in 1965, Hessel and Oliner clearly distinguished two types of Wood's anomalies: square-root-type singularities at the scattering thresholds and those that can be attributed to resonances~\cite{Hessel1965}. Whereas the first type of singularity arises from the square root's double-valued nature and its branch point at the scattering thresholds, resonances, also referred to as quasinormal modes (QNMs) or leaky modes, are solutions to the source free Maxwell's equation characterized by their nonzero imaginary parts modeling the dissipation of energy over time. They commented that the derivative of the square root results in an infinite slope, which they anticipated would boost resonance effects.
Although each of these aspects is well established in nanophotonics, it is the combination of resonances and diffraction that currently attracts interest and has led to the development of metasurfaces with Mie resonances, plasmon resonances, and optical bound states in the continuum (BICs) \cite{Babicheva2024,Ding_DeGruyter2018,Koshelev_PRL_2018,Abajo2007,GarciaVidal2010,Kravets2018,Wang2018}. 
In fact, it has been proposed to exploit BICs close to a scattering threshold to maximize the sensitivity of refractive index sensing~\cite{Maksimov2022PRA}.

Recently, a mathematical model for the entries of the scattering matrix in the vicinity of scattering thresholds has been presented~\cite{Wojcik2021PRL}. Whereas the model accurately reproduces the square root behavior, it does not include resonances.
In contrast, two established formalisms, known as QNM expansion~\cite{Lalanne_QNMReview_2018,Franke_PRL_2019,Nicolet_2023} and resonant state expansion~\cite{Muljarov_EPL_2010}, respectively, have been put forward to describe resonant behavior. However, their application to grating systems is known to be subtle. Indeed, for such formalisms, it has been pointed out that algorithms identify a plethora of additional resonances without a direct physical interpretation, but that account for the discontinuity of the derivative at the scattering threshold~\cite{Doost2013PRA,Gras2019OptLett}. These numerical resonances, as explicitly stated in ref.~\cite{Doost2013PRA}, discretize an integral. 
With an arbitrary black-box solver for Maxwell's equation at hand, distinguishing physical resonances from those representing the branch cut is a non-trivial task. An alternative approach that explicitly evaluates resonances in the vicinity of scattering thresholds describes the optical response as a function of a wave vector component instead of frequency~\cite{akimov2011optical,Armitage2014PRA}. The corresponding transformed space does not contain branch points.

\begin{figure*}
\includegraphics{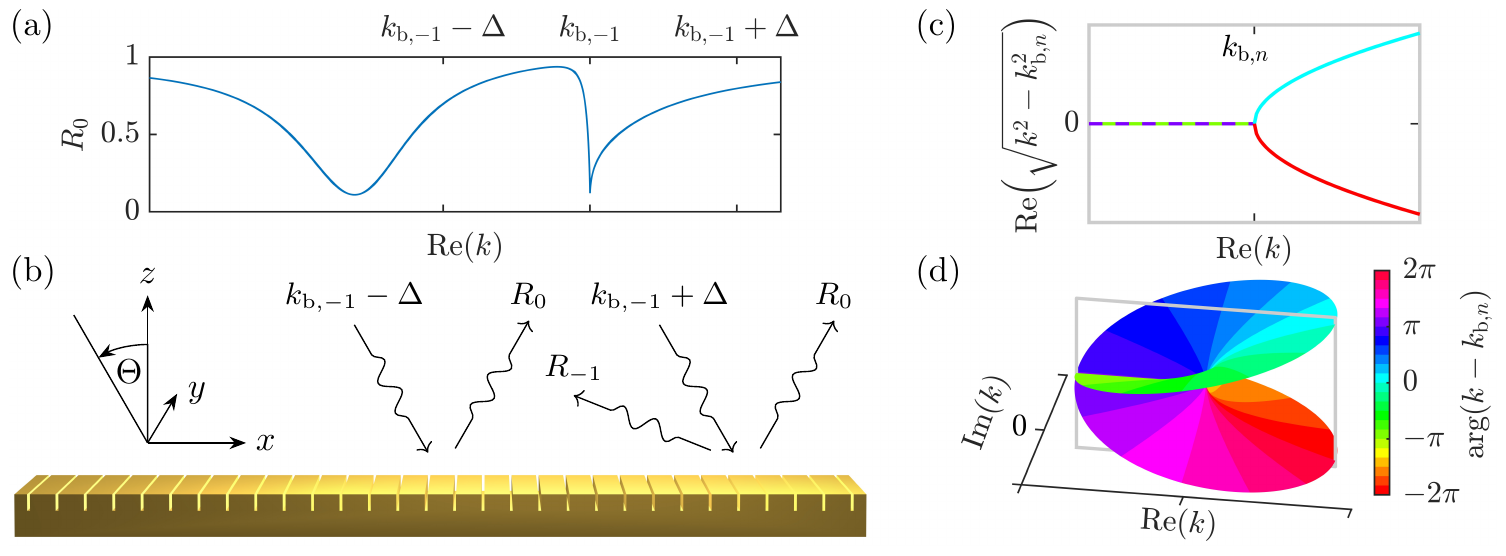}
\caption{\label{fig1} 
Outline of the approach following the panels in indicated order: The specular reflection spectrum $R_0(k)$ exhibits a branch point where the $z$-component $k_{z,1}$ of the first order reflection $R_1$ transitions from purely imaginary (evanescent) to purely real (propagating). Samples of the the zero order Fourier coefficient $f_j$ are evaluated at various real valued $k$ in the vicinity of the branch point. The transformation, used to map the sampling points $k_j$ to $\tilde k_j$, locally provides a map of the Riemann surface. The AAA algorithm uses a subset $I$ of the samples to construct a rational approximation in barycentric form interpolating the $f_j$ with $j\in I$ and approximating all other data points in a least square sense. The rational function provides an analytic continuation of the data points to the full $\tilde k$-space and its poles correspond to resonances of the physical system. In the physical domain a branch cut is introduced, that cuts the Riemann surface in two distinct sheets, each defined over the entire complex plane. We choose a branch cut parallel to the negative imaginary axis, which in the example at hand approximately corresponds to the line $\mathrm{Im}\bigl(\tilde k\bigr)+\mathrm{Re}\bigl(\tilde k\bigr)=0$. We refer to poles as hidden resonances when they do not reside on the Riemann sheet that contains the real axis, as their influence on the spectrum is substantially concealed by the branch point.
}
\end{figure*}

In this work, we use such a transformed space to access the relevant resonances on different Riemann sheets of periodic non-Hermitian systems and reconstruct the optical response across several scattering thresholds with a multi-valued rational approximation. The results enable an instructive interpretation of the interaction between resonances and square-root-type singularities. Using the example of a plasmonic line grating, upon changing a design parameter, we observe resonances vanish behind a branch point, and others appear in the spectrum that had been hidden previously. In this way, a distinctive feature seemingly unrelated to resonances is accurately described with inherent properties of the system.

\section{Multi-Valued Rational Approximation}

The AAA algorithm~\cite{Nakatsukasa_2018_AAA} has proven to be an efficient method to reconstruct optical response functions with rational functions $r(k)$, $k\in\mathbb{C}$~\cite{AAACNO}. An outline of the approach introduced here is provided in Fig.~\ref{fig1}. We are interested in the response of a periodic system as a function of the incident light wave number $k$. We expect sudden drops in intensity due to Wood's anomalies, that is, resonance effects and square-root-type singularities, as well as their interactions. The spectrum in the last panel of Fig.~\ref{fig1} shows two pronounced dips. 
The broad minimum at lower wave numbers corresponds to a plasmon resonance of the system. At higher wave numbers, there is a sharp feature with a discontinuous derivative. This feature at the point labeled $k_{\mathrm{b},1}$ corresponds to the appearance of the first-order reflection $R_{1}$. A rational function $r(k)$ can approximate the plasmon resonance with a single pole, but not the second feature. As an alternative to the wave number $k$, the $z$-component of the wave vector can be used to parameterize the response~\cite{akimov2011optical}. In this transformed representation, the specular reflection can be fully approximated with a low-order rational function. Subsequently, $r(k)$ is obtained by a coordinate transformation.

Given a structure periodic in $x$ and $y$, the $z$-component $k_z$ of the diffracted light wave vector is constrained by the condition 
\begin{equation} \label{eq:kz}
    k_z = \sqrt{k^2-\left|\mathbf{k}_\parallel+\mathbf{G}\right|^2}\,,
\end{equation}
where $\mathbf{G}$ is a reciprocal lattice vector and $\mathbf{k}_\parallel$ the components of the wave vector parallel to the periodic structure. Each diffraction order corresponds to a specific $\mathbf{G}$ and will contribute to the propagating spectrum if the term inside the square root is positive. For $k^2-\left|\mathbf{k}_\parallel+\mathbf{G}\right|^2<0$ the normal component $k_z$ is purely imaginary, which results in an evanescent diffraction order. Please refer to the first panel of Fig.~\ref{fig1}, which illustrates the emergence of the first diffraction order of a line grating. We are interested in the branch points $k_{\mathrm{b},n}$ with $n\in \mathcal{N}\subset \mathbb{Z}$ that mark the emergence of new diffraction orders in the propagating spectrum. They are solutions of $k_{\mathrm{b},n}-\left|\mathbf{k}_\parallel+\mathbf{G}\right|=0$. Using $N = |\mathcal{N}|$ of these solutions, we define the coordinate transform
\begin{equation} \label{eq:ktilde}
    \tilde{k} = \frac{1}{N}\sum_{n\in\mathcal{N}}\sqrt{k^2 - k_{\mathrm{b},n}^2}\,.
\end{equation}
This transformation asymptotically approximates $k$ and shows locally, around its branch points $k_{\mathrm{b},n}$, the square root behavior of $k_z$. The choice of the transformation is not unique. The requirement is a map of the full Riemann surface near the branch points. Alternatives could be a sum over different reciprocal lattice vectors $\mathbf{G}$ in Eq.~\eqref{eq:kz} and a Schwarz-Christoffel mapping~\cite{Driscoll_Trefethen_2002}.

The square roots in the definition of the complex variable $\tilde{k}$ render it a multi-valued function of $k$ and the same holds true for the function $r_{\tilde k}(k) = r\bigl(\tilde k(k)\bigr)$. Defining branch cuts parallel to the negative imaginary axis, we term singularities which are not located on the Riemann sheet containing the real axis hidden resonances. The need to define a branch cut is illustrated in the last two panels of Fig.~\ref{fig1}. The black line across the $\tilde k$-plane in the fifth panel defines the branch cut in the sixth panel, which cuts the Riemann surface in two sheets both defined over the entire $k$-plane. Approaching the cut from different sides results in different function values. A rational function of $\tilde{k}$ can have poles on any Riemann sheet, and therefore, introducing a branch cut hides information. Unlike ref.~\cite{Smith1992}, where resonances are evaluated directly on a four-sheeted Riemann surface, we solve the resonance problem here in a transformed space using the AAA algorithm \cite{Nakatsukasa_2018_AAA}. As indicated in the fourth panel of Fig.~\ref{fig1}, the algorithm takes $M$ sampling points and greedily adds terms to a rational approximation $r\bigl(\tilde k\bigr)$ of the optical response function $R\bigl(\tilde k\bigr)$
\begin{equation} \label{eq:bary}
    R\bigl(\tilde k\bigr) \approx r\bigl(\tilde{k}\bigr) = \left. \sum_{j\in I}\frac{w_j R_j}{\tilde{k} - \tilde{k}_j}\middle/\sum_{j\in I} \frac{w_j}{\tilde{k} - \tilde{k}_j} \right.\,.
\end{equation}
By definition, the barycentric form in Eq.~\eqref{eq:bary} interpolates the function values $R_j =R\bigl(\tilde k_j\bigr)$ at the subset of sampling points $\tilde k_j$ with $j\in I \subset \{1,\dots,M\}$. In each iteration, the weights $w_j$ under the constraint $\sum_{j\in I} |w_j|^2 = 1$ are chosen to minimize the error $\bigl|r\bigl(\tilde k_i\bigr)-R\bigl(\tilde k_i\bigr)\bigr|^2$ at samples $\tilde k_i$ with $i\notin I$.

The resulting poles can be identified with the resonances of the physical system~\cite{AAACNO} and are mapped back to the $2N$ Riemann sheets introduced with $\tilde{k}$. This back transformation is crucial to interpret their impact on the physical response near branch points, and therefore we use the term {\it multi-valued approach}. 

\section{Application}

\begin{figure}
\includegraphics{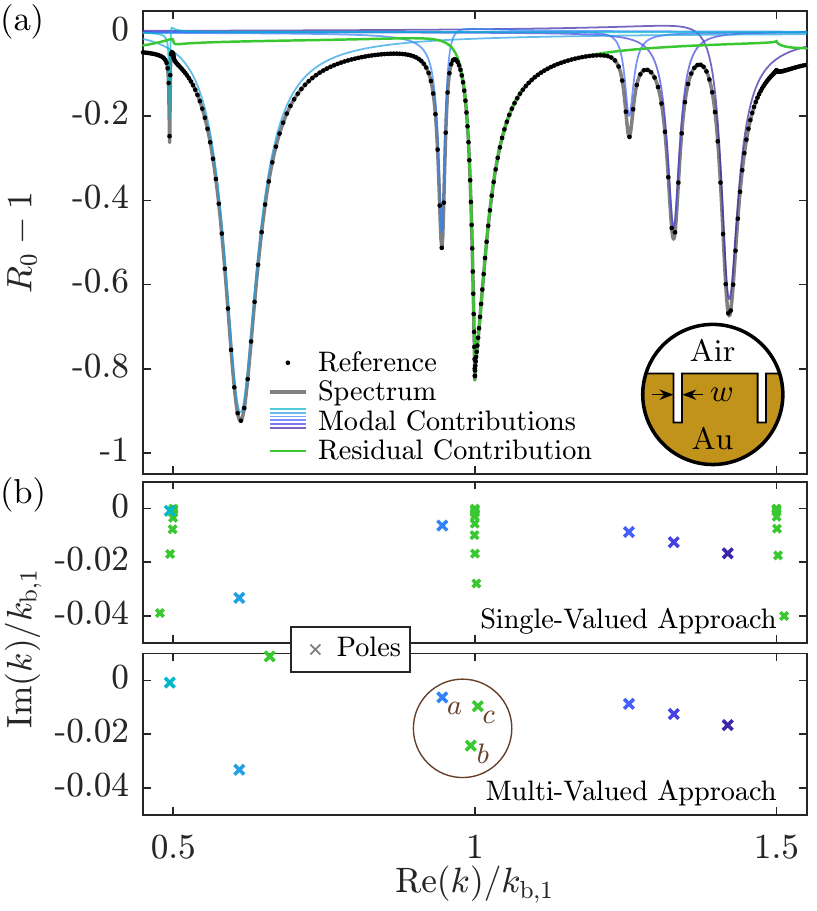}
\caption{\label{fig2}
The specular reflection spectrum $R_0(k)$ across the three branch points $k_{\mathrm{b},1} = 0.5 \, k_{\mathrm{b},2}$, $k_{\mathrm{b},2} = \qty{1.074e7}{\per \meter}$ and $k_{\mathrm{b},3} = 1.5 \, k_{\mathrm{b},2}$. The gold grating, sketched in the circular inset, is illuminated by a TM-polarized plane wave at an incidence angle of $\Theta=30^\circ$. (a) The approximated spectrum together with reference points, a decomposition into six modal contributions and the residual contribution, which remains after subtracting the modal contributions from the full spectrum. (b) Resonances of the system: six resonances (blue crosses) are found at identical positions with and without the coordinate transformation $\tilde{k}$. In the single-valued approach (top), additional resonances cluster at wave numbers close to the branch points. Conversely, the multi-valued approach (bottom) reveals three additional resonances. The labels of the resonances within the circle refer to Fig.~\ref{fig3}.}
\end{figure}

We investigate the interaction between resonances and scattering thresholds using the example of a line grating sketched in the inset of Fig.~\ref{fig2}(a) that supports pronounced grating anomalies and is immersed in a medium with refractive index $n = 1.3$~\cite{Gras2019OptLett}.
A Drude-Lorentz model specifies the relative permittivity of gold $\varepsilon(\omega) = \varepsilon_0 - \varepsilon_0\omega_p^2 (\omega^2 + i \omega \gamma)^{-1}$, with the vacuum permittivity $\varepsilon_0$ and the plasma frequency $\omega_p = \qty{1.26e16}{\radian \per \second}$, the damping factor $\gamma = 0.0112 \, \omega_p$, and the angular frequency $\omega = c k$ that is related to the wave number $k$ through the speed of light $c$. Working in the optical regime, we can equate the permeability $\mu(\omega) = \mu_0$ with that in the vacuum. The grating period $p$, which defines the one-dimensional lattice vector $G = 2\pi n/p$ with $n\in\mathbb{Z}$, is \qty{600}{\nano \meter}, and the depth of the rectangular grooves is \qty{350}{\nano \meter}. The width $w$ of \qty{51}{\nano \meter} maximizes the interaction between the square-root-type singularity and resonances at the angle of incidence $\Theta = 30^\circ$. For the definition of the angle, we refer to the first panel of Fig.~\ref{fig1}.

Samples of the optical response at real-valued frequencies are the only prerequisites for the proposed method. Here, we rely on the finite element method (FEM) and its implementation in the JCMsuite software package~\cite{Pomplum_NanoopticFEM_2007} to solve the second-order Maxwell equation
\begin{equation}
	\nabla \times \mu^{-1} 
	\nabla \times \mathbf{E} -
	\omega^2\epsilon \mathbf{E}  = 
	i\omega\mathbf{J}\,, \label{eq:Maxwell}
\end{equation}
with the electric current density $\mathbf{J}$  and the electric field strength $\mathbf{E}$. With plane wave illumination, the upward directed Fourier spectrum of $\mathbf{E}$ yields the specular reflection $R_0$, which is the absolute value squared of the zero-order coefficient. 

The condition $k_{\mathrm{b},n}-\left|k_{\mathrm{b},n} \mathrm{sin}\Theta+2\pi n / p \right|=0$ for the scattering thresholds $k_{\mathrm{b},n}$ follows from Eq.~\eqref{eq:kz}. 
Figure~\ref{fig2}(a) shows the specular reflection $R_0(k)-1$ in a range that spans the three scattering thresholds $k_{\mathrm{b},1}$, $k_{\mathrm{b},2}$, and $k_{\mathrm{b},3}$, as obtained by a usual driven diffraction calculation. 
In addition, the modal contributions of six dominant resonances in the spectrum are displayed. 
This modal expansion of $R_0$, which is quadratic in the electric field, is based on the residues of the Fourier coefficient and evaluations of its rational approximation at the complex conjugated resonance frequencies following~\cite{Betz_2022}. Subtracting the sum of the six modal contributions from the full spectrum gives the residual contribution that exhibits the square-root-type Wood's anomalies.

Using Eq.~\eqref{eq:ktilde} to incorporate the three relevant branch points into the rational approximation, the AAA algorithm requires 80 support points to reduce the relative error of the reconstructed spectrum with respect to the reference points below $10^{-6}$, which is close to the accuracy of the FEM simulations. Without the coordinate transformation, the error is three orders of magnitude larger. For details on the sampling strategy and the convergence, we refer to the associated data publication~\cite{Betz_Zenodo_BranchPoints}.

The lower part of Fig.~\ref{fig2}(b) shows the resonances $k_n$ of a rational approximation without coordinate transformation. In this case, the algorithm clusters many additional resonances along characteristic lines close to the scattering thresholds. The contributions of these resonances approximate integrals along branch cuts, as discussed in ref.~\cite{Doost2013PRA} and give rise to the discontinuities in the derivative. We will show that the feature at $k_{\mathrm{b},1}$ that particularly catches the eye is the result of two hidden resonances on different Riemann sheets. 

The lower section of Fig.~\ref{fig2}(b) shows resonances $k_{\mathrm{p},m}$ derived from the poles of the rational approximation $r\bigl(\tilde{k}\bigr)$ that results from an application of the AAA algorithm in the transformed space. Given the poles $\tilde{k}_{\mathrm{p},m}$ of $r\bigl(\tilde{k}\bigr)$ and the single-valued definition of the square root with a branch cut parallel to the negative imaginary axis, the resonances marked with green crosses do not solve the radical equation 
\begin{equation} \label{eq:kn}
    \tilde{k}_{\mathrm{p},m} = \frac{1}{N}\sum_{n=1}^N\sqrt{k_{\mathrm{p},m}^2 - k_{\mathrm{b},n}^2}
\end{equation}
that follows from Eq.~\eqref{eq:ktilde}. These hidden resonances do not appear on the sheet that contains the physical signal, but they do impact the response. It should be noted that the pole above the real axis enters the physical sheet at $k_{b,1}$ with a negative imaginary part if a parameter is changed accordingly. For a better understanding of hidden resonances, we will study in the following how the two resonances near the branch point $k_{\mathrm{b},1}$ labeled $b$ and $c$ behave if the width of the grooves $w$ is varied. 

\begin{figure}
\includegraphics{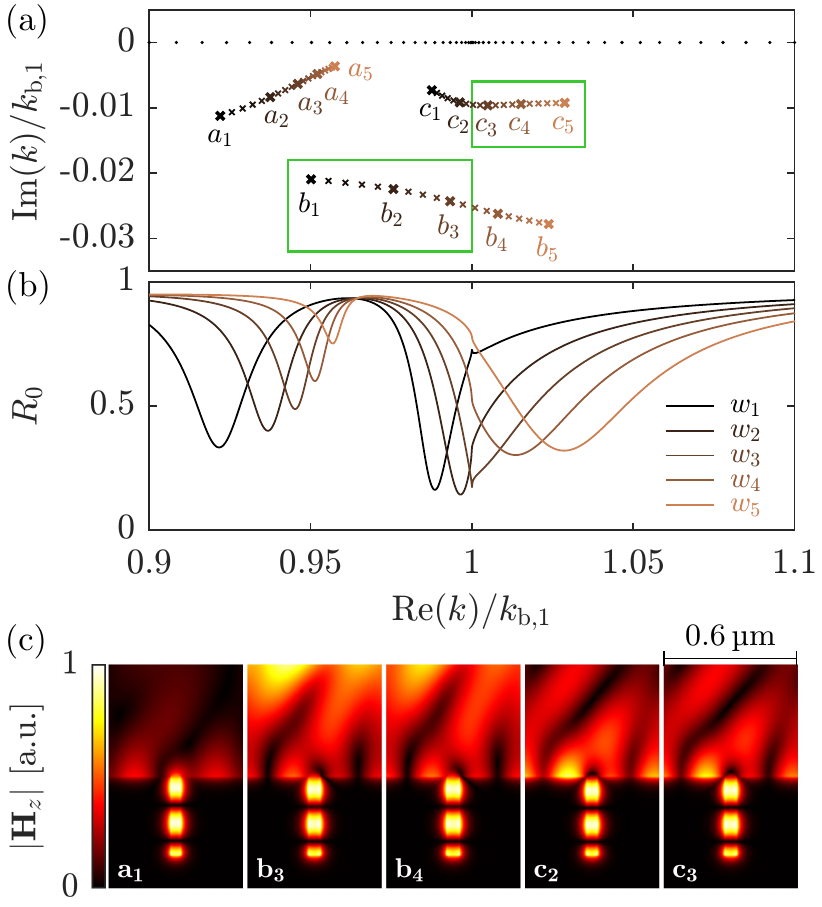}
\caption{\label{fig3}
Resonances passing a branch point. (a) The three resonances closest to the 50 support points (black dots) are displayed as a function of the groove width $w$, which is varied in steps of \qty{1}{\nano\meter} from $w_1 = \qty{42}{\nano \meter}$ to $w_5 = \qty{60}{\nano \meter}$. The resonances inside the green boxes are not directly connected to the real axis above them, i.e., they are hidden resonances. (b) Specular reflection spectra corresponding to the labeled subset of resonances. (c) The magnetic field patterns corresponding to the three relevant resonances at selected widths.
}
\end{figure}

The trajectories of three resonances labeled $a$, $b$, and $c$ when varying $w$ are presented in Fig.~\ref{fig3}(a). The hidden resonances that require both Riemann sheets to solve Eq.~\eqref{eq:kn} are highlighted with green boxes. A comparison with the spectra at selected widths shown in Fig.~\ref{fig3}(b) reveals that when $w$ is increased, the spectral feature linked to the resonance $c$ approaches the branch point and is eventually replaced by the broader spectral feature of $b$, which had previously been hidden. Hence, instead of using a large number of numerical resonances to represent the branch cut, few resonances fully explain the physical response. In the current scenario, only two resonances are needed. A special case arises for the width $w_3$, as both $b_3$ and $c_3$ are hidden resonances, and the signal is highly asymmetric, with one flank steeper than the other. Although the first derivative of the physical response is highly discontinuous, Fig.~\ref{fig3}(c) shows that the field patterns of the resonances continuously vary as $w$ changes. This observation emphasizes that Fig.~\ref{fig3}(b) does not show the signal of a single pole crossing the branch point but rather two distinct resonances. 

\begin{figure}
\includegraphics{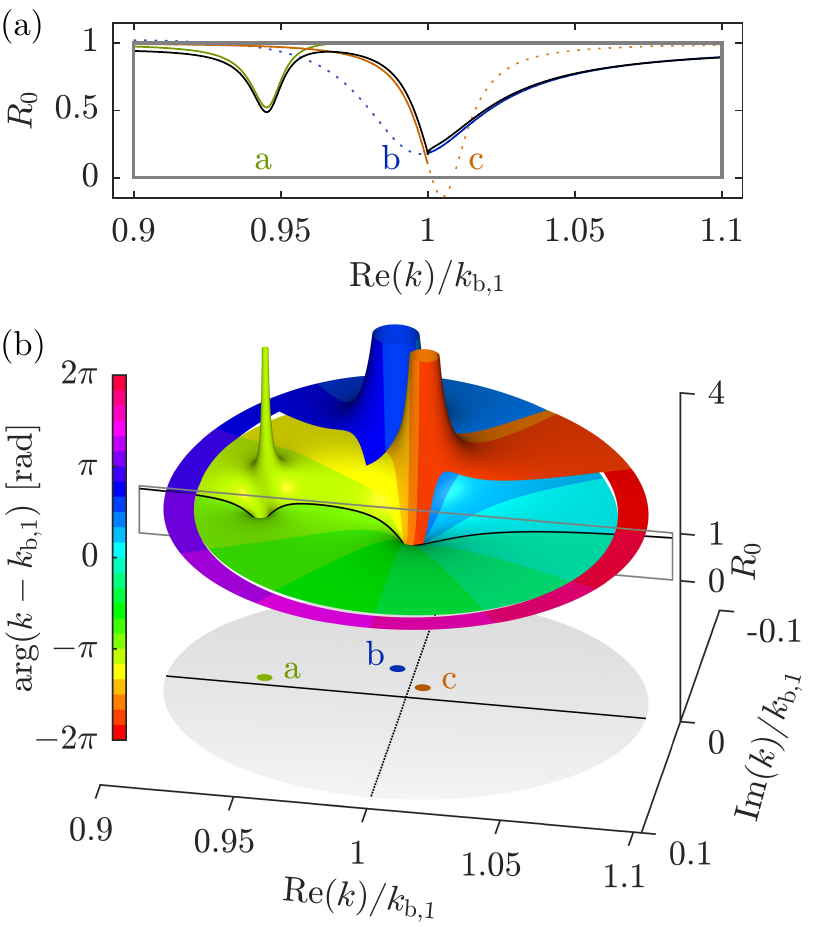}
\caption{\label{fig4}
Multi-valued specular reflection. (a) The contributions of the resonances $b_3$ and $c_3$ explain the shape of the specular reflection (black line) at the respective sides of the branch point. (b) The Riemann surface of the double-valued specular reflection. The positions of the resonances are projected to the complex plane in the bottom part of the graph. The colors refer to the argument of $k-k_{\mathrm{b},1}$ that ranges from $-2\pi$ to $2 \pi$ to uniquely define every point of the double-valued function.
}
\end{figure}

We conclude the investigation of resonances in the vicinity of branch points with a picture visualizing their effect on the optical response measured along the real axis. We have observed in Fig.~\ref{fig3} that the resonances $b$ and $c$ are responsible for the asymmetric shape of the spectral feature. This finding is confirmed in Fig.~\ref{fig4}(a). 
At smaller wave numbers, the flank of the feature follows the resonance contribution of $c$ even though the resonance itself is shifted to higher wave numbers, where the shape of the feature can be explained with the contribution of $b$. The complete Riemann surface of the specular reflection in Fig.~\ref{fig4}(b), with colors visualizing the phase $\varphi = \mathrm{arg}(k-k_{\mathrm{b},1})$, explains why the peak at the point $b$ mainly influences the spectrum at wave numbers larger than $k_{b,1}$, even though $\mathrm{Re}(b)<k_{b,1}$. Since the angles of the physical signal are $-\pi$ and zero, the position $b$ with $\varphi>\pi/2$ hardly affects the signal in $\varphi=-\pi$, where the resonance at $c$ contributes dominantly.

\section{Conclusion}

We demonstrated that multi-valued rational approximations accurately describe resonances near scattering thresholds. The resulting low-order model explains the characteristic sharp features with resonances on different Riemann sheets.  We show how these hidden resonances can strongly enhance sharp features in the scattering response. 
The model accurately describes the discontinuity of the first derivative of the optical response function. 
Therefore, it is possible to reconstruct complex spectra with different types of singularity, as presented in Fig.~\ref{fig2}(a), with a single rational approximation.

The proposed method is particularly easy to implement, as it only requires samples of the optical response at real-valued frequencies that can be collected numerically or experimentally. Suitable numerical methods are, among others, the rigorous coupled wave analysis (RCWA), the boundary element method (BEM), and, as in the example at hand, the finite element method (FEM).

We envision a wide use of our method in a variety of fields, including optics and electromagnetism, as well as acoustics. Engineering of response functions by hybridization of square-root-type singularities and resonances is a very common motif in the design of plasmonic and dielectric structures for wavefront control, fluorescence control, sensing, and nonlinear optics.  In the plasmonic domain this includes, for instance, the field of surface lattice resonances, i.e., high quality collective plasmon resonances in periodic metal nanoparticle lattices that have been developed for fluorescence control,  surface enhanced Raman scattering and lasing \cite{Kravets2018,odom2022,lozano2013}.  In the field of dielectric metasurfaces and photonic crystals there is a strong interest in Fano resonances and quasi-bound states in the continuum \cite{Limonov_NatPhot_2017,koshelev2018}, in which Mie resonances~\cite{koshelev2021}, square-root-type singularities and guided mode resonances interplay. These structures are pursued for their high quality factors and strong local field confinement, with demonstrated applications in IR spectroscopy \cite{tittl2018},  nonlinear optics and high harmonic generation \cite{liu2018,koshelev2019}, and nonlocal metasurfaces for wavefront control \cite{overvig2021}. Our method provides an efficient and intuitive approach to attribute the response of these systems to just a few resonant contributions, including hidden resonances. An interesting prospect is to track these resonances to enable optimization schemes based on their positions~\cite{Binkowski_CommunPhys_2022}, even when they are partially obscured by the branch point. 

\emph{Data availability.}---Supplementary data tables, details on the convergence and source code for the numerical experiments of this work can be found in the open-access data publication~\cite{Betz_Zenodo_BranchPoints}.

\emph{Acknowledgments.}---FB, FB, LZ, and SB acknowledge funding
by the Deutsche Forschungsgemeinschaft (DFG, German Research Foundation) 
under Germany's Excellence Strategy - The Berlin Mathematics Research
Center MATH+ (EXC-2046/1, project 390685689) and
by the German Federal Ministry of Education and Research
(BMBF Forschungscampus MODAL, project 05M20ZBM). 
JDF and CR acknowledge ﬁnancial support by the Helmholtz Association in the framework of the innovation platform “Solar TAP”. 
The work of NF and AFK is part of the Dutch Research Council (NWO) and was performed at the research institutes AMOLF and ARCNL. 
ARCNL (Advanced Research Center for Nanolithography) is a public-private partnership between the University of Amsterdam, Vrije Universiteit Amsterdam, University of Groningen, NWO, and the semiconductor equipment manufacturer ASML.

\bibliography{bibliography}

\end{document}